\documentclass[aps,prd,showpacs,twocolumn]{revtex4-1}

\usepackage{amsmath,amssymb}
\usepackage{graphicx}

\usepackage{bm}

\usepackage{color}
\usepackage{subfigure} 

\newcommand*{\mysim}{\mathord{\sim}}

\begin{document}

\title{Efficient Estimation of Highly Structured Posteriors of
  Gravitational-Wave Signals with Markov-Chain Monte Carlo}

\author{Benjamin Farr}
  \affiliation{Center for Interdisciplinary Exploration and Research in
    Astrophysics (CIERA) \& Dept.\ of Physics and Astronomy, Northwestern
    University, 2145 Sheridan Rd, Evanston, IL 60208, USA}

\author{Vicky Kalogera}
  \affiliation{Center for Interdisciplinary Exploration and Research in
    Astrophysics (CIERA) \& Dept.\ of Physics and Astronomy, Northwestern
    University, 2145 Sheridan Rd, Evanston, IL 60208, USA}

  \author{Erik Luijten} 
  \affiliation{Center for Interdisciplinary Exploration and Research in
    Astrophysics (CIERA) \& Dept.\ of Materials Science and Engineering
    \& Dept.\ of Engineering Sciences and Applied Mathematics,
    Northwestern University, 2220 Campus Dr, Evanston, Illinois 60208,
    USA}

\begin{abstract}
    We introduce a new Markov-Chain Monte Carlo (MCMC) approach designed for
    efficient sampling of highly correlated and multimodal posteriors.
    Parallel tempering, though effective, is a costly technique for sampling
    such posteriors.  Our approach minimizes the use of parallel tempering,
    only using it for a short time to tune a new jump proposal.  For complex
    posteriors we find efficiency improvements up to a factor of $\mysim 13$.
    The estimation of parameters of gravitational-wave signals measured by
    ground-based detectors is currently done through Bayesian inference with
    MCMC one of the leading sampling methods.  Posteriors for these signals are
    typically multimodal with strong non-linear correlations, making sampling
    difficult.  As we enter the advanced-detector era, improved sensitivities
    and wider bandwidths will drastically increase the computational cost of
    analyses, demanding more efficient search algorithms to meet these
    challenges.
\end{abstract}


\maketitle

\section{Introduction}

In the coming years, the detectors of the Laser Interferometer
Gravitational-Wave Observatory (LIGO) and Virgo Collaboration (LVC) will come
online following a multi-year endeavor to upgrade the instruments.  This
so-called ``advanced-detector era'' will ultimately bring a projected factor of
10 increase in range, and a broadened band of sensitivity reaching down to 10
Hz from the previous era's lower limit of 40 Hz~\citep{Aasi:2013wya,adLIGO}.
This additional sensitivity at lower frequencies means that gravitational waves
(GWs) from compact binaries will become detectable at even earlier times before
merger than was previously possible.  This increases measured waveform lengths
by a factor of $\mysim 40$, increasing the duration of the longest measurable
signals from tens of seconds to tens of minutes.

To estimate the parameters of a GW source, the LVC parameter estimation (PE)
algorithms (found in the \emph{LALInference} software
package~\citep{S6PE,PEmethods}) compute $\mysim 10^7$--$10^8$ model waveforms
that are compared to the interferometric data.  Because the generation of these
waveforms constitutes the computational bottleneck of the analysis, the longer
waveforms required for advanced LVC parameter estimation will increase analysis
run times by up to a factor of $\mysim 50$.  PE analyses required several hours
to several days to analyze a GW candidate at the end of the last science run,
when a GW entered the band of sensitivity at 40 Hz~\citep{S6PE}; without further
optimization the analysis of individual GW candidates in the advanced detector
era will be prohibitively long. Improvements to both model waveforms and PE
algorithms will likely be necessary for us to be prepared for parameter
estimation in the advanced detector era. This work addresses inefficiencies of
the PE methods currently employed, particularly focusing on Markov-Chain Monte
Carlo (MCMC) methods.  We propose a new analysis method that promotes more
effective exploration of the parameter space and significantly reduces the
total number of waveforms that need to be generated.  We emphasize that
although this algorithm was developed to aid in the parameter estimation of GW
sources, the techniques are not problem-specific, and can potentially be
applied to other MCMC algorithms to increase the efficiency of estimating
highly structured posteriors.

In Section~\ref{sec:models} we give a brief introduction to the noise and the
signal models for GWs from binary inspirals. Section~\ref{sec:est} outlines the
MCMC methods employed by \emph{LALInference} in the LVC's last science run.
Section~\ref{sec:newMCMC} describes the new MCMC strategy that we have
developed to increase sampling efficiency.

\section{Signal and Noise Models}
\label{sec:models}
The Bayesian PE algorithms used to analyze LVC data depend on models for both
the noise and the signal.  To provide context, here we briefly discuss those
models. The most accurate models for GWs produced by compact binary systems are
those generated by simulations that numerically solve the full non-linear
differential equations of general relativity.  However, this approach is
computationally far too expensive to be used for PE analyses. Therefore, in
lieu of numerical waveforms, PE algorithms rely on approximate methods such as
post-Newtonian expansion~\cite{PNreview} or the effective-one-body
formalism~\cite{EOB} to generate model waveforms for a given set of physical
parameters~$\boldsymbol{\theta}$.

The GWs produced by a quasi-circular compact binary system of masses $m_1$
and~$m_2$ are parametrized by fifteen parameters~\cite{PNreview},
\begin{equation}
  \label{eqn:params}
  \boldsymbol{\theta} = \left\{ \mathcal{M}_c, q, \boldsymbol{S}_1,
    \boldsymbol{S}_2, \iota, D_L, \psi, \alpha, \delta, t_c,
    \phi_c\right\} \;,
\end{equation}
where $\mathcal{M}_c = (m_1m_2)^{3/5}/(m_1+m_2)^{1/5}$ is the chirp mass,
$q=m_2/m_1$ the asymmetric mass ratio defined such that $0<q\leq1$,
$\boldsymbol{S}_i$ the spin vector of $i^\text{th}$ binary component, $\iota$
the inclination of the orbital plane relative to the observer's line of sight,
$D_L$ the luminosity distance, $\psi$ the polarization angle, $\alpha$ the
right ascension, $\delta$ the declination, $t_c$ the time of coalescence, and
$\phi_c$ the phase at coalescence.  For the purposes of this work, we will focus
only on mergers of non-spinning compact objects where $|\boldsymbol{S}_1| =
|\boldsymbol{S}_2| = 0$, reducing the parameter space to nine dimensions.

The current noise model used for PE analyses assumes the noise to be stationary
and Gaussian, with a power spectral density that is estimated via Welch's
method~\cite{PSDest} near the time of interest (i.e., trigger
time)~\cite{S6PE}.  However, real detector noise is often non-stationary and
non-Gaussian, with occasional glitches and non-stationarities not currently
accounted for in the noise model~\citep{S5glitch} that can potentially bias
parameter estimates.  More sophisticated noise modeling is outside the scope of
this work, but remains an area of active research~\cite{noise1,noise2}.

\section{MCMC Techniques and Parameter Estimation}
\label{sec:est}

The \emph{posterior} probability $p(\boldsymbol{\theta}|d)$ of the parameter
set~$\boldsymbol{\theta}$ given the data~$d$ is calculated according to Bayes'
theorem, a framework for updating prior information $\pi(\boldsymbol{\theta})$
based on newly measured data,
\begin{equation}
  \label{eqn:posterior}
  p(\boldsymbol{\theta}|d) =
  \frac{\pi(\boldsymbol{\theta})\mathcal{L}(\boldsymbol{\theta})}{p(d)} \;,
\end{equation}
where the \emph{likelihood}
$\mathcal{L}(\boldsymbol{\theta})=p(d|\boldsymbol{\theta})$ is the probability
of measuring the data $d$ given the parameter set $\boldsymbol{\theta}$, and
$p(d)$ is the \emph{marginal likelihood}.  The likelihood function for a
detector network is given by the product of individual detector
likelihoods~\citep{singleIFO},
\begin{equation}
  \label{eqn:likelihood}
  \mathcal{L}(\boldsymbol{\theta}) \propto \prod_i \exp\left[ -2
    \int_0^\infty \frac{|\widetilde{d_i}(f) - \widetilde{h_i}(f,
      \boldsymbol{\theta})|^2}{S_{n,i}(f)}df\right] \;,
\end{equation}
where $\widetilde{d_i}(f)$, $\widetilde{h_i}(f,\boldsymbol{\theta})$, and
$S_{n,i}(f)$ are the $i^\text{th}$ detector's data, modeled signal, and
one-sided noise power spectral density, respectively, in the frequency domain.

To define our formalism and notation we briefly summarize the basics of
Bayesian analysis and MCMC methodology.  The posterior
distributions~\eqref{eqn:posterior} of compact binary GW signals in the LVC are
typically estimated using multiple sampling algorithms; nested
sampling~\citep{NS}, MultiNest~\citep{MN}, and MCMC~\citep{MCMC1,MCMC2} have
all proven to be effective sampling techniques.  Here we introduce several
improvements aimed at the MCMC approach, but some of the proposed techniques
(in particular the tuned jump proposal outlined in Sec.~\ref{subsec:optics})
may improve the efficiency of other sampling techniques as well.

MCMC methods produce samples at a density proportional to that of the target
posterior distribution by constructing a Markov chain whose equilibrium
distribution is proportional to the posterior distribution.  Our MCMC
implementation uses the Metropolis--Hastings
algorithm~\citep{metropolis53,hastings}, which requires a proposal density
$Q(\boldsymbol{\theta}'|\boldsymbol{\theta})$ to generate a new sample
$\boldsymbol{\theta}'$ given the current sample $\boldsymbol{\theta}$.  Such a
proposal is accepted with a probability $r_s = \text{min}(1, \alpha)$, where
\begin{equation}
  \label{eqn:acceptance}
  \alpha = \frac{Q(\boldsymbol{\theta}|\boldsymbol{\theta}')
    p(\boldsymbol{\theta}'|d)}{Q(\boldsymbol{\theta}'|\boldsymbol{\theta})
    p(\boldsymbol{\theta}|d)} \;.
\end{equation}
If accepted, $\boldsymbol{\theta}'$ is added to the chain, otherwise
$\boldsymbol{\theta}$ is repeated.

Chains are typically started at a random location in parameter space, requiring
some number of iterations before dependence on this location is lost.  The
samples collected during this \emph{burn-in} period are necessary, but useless,
as they are usually discarded when estimating the posterior.  Furthermore
adjacent samples in the chain are typically correlated to some degree,
requiring the chain to be thinned by its integrated autocorrelation time (ACT).
We refer to the samples remaining after discarding burn-in and thinning by the
ACT as the \emph{effective samples}.

The efficiency of the Metropolis--Hastings algorithm is largely dependent on
the choice of proposal density, since that is what governs the acceptance rates
and ACTs.  The most commonly used proposal density is a Gaussian centered on
$\boldsymbol{\theta}$.  The width of this Gaussian for each parameter will
affect the acceptance rate of the proposal.  Widths that are too large will
cause low acceptance rates, whereas widths that are too small will lead to
strongly correlated samples and large ACTs.  For the idealized case of a
posterior on $\mathbb{R}^d$ composed of independent and identically distributed
components such that $p(\theta_1,\theta_2,\ldots,\theta_d) =
f(\theta_1)f(\theta_2)\ldots f(\theta_d)$, where $f$ is a one-dimensional (1D)
smooth density, it can be shown that the optimal acceptance rate is
approximately $0.234$~\citep{adaptMCMC}.  This value, applicable only to the
local Gaussian jump proposal, provides the optimum balance between acceptance
rate and ACT.  In principle, proposals can achieve arbitrarily high acceptance
rates and yet produce uncorrelated samples---as shown in the context of MCMC
schemes in statistical mechanics~\cite{geomc,liu05a,sinkovits12b}.
Nevertheless, we find that for typical situations in GW data analysis,
targeting an acceptance rate of~$0.234$ allows for relatively consistent ACTs
for all posteriors.  Therefore, during the burn-in period we scale the 1D
Gaussian widths of all proposal densities by a factor that decays with the
fifth root of the iteration number to approximately achieve this acceptance
rate.

Gaussian jump proposals are typically sufficient for unimodal posteriors and
spaces without strong correlations between parameters.  However, there are many
situations where strong parameter correlations exist and/or multiple isolated
modes appear spread across the multi-dimensional parameter space.  When
parameters are strongly correlated, the ideal jumps would be along these
correlations.  This makes the 1D jumps in the model parameters very
inefficient.  Furthermore to sample between isolated modes, a chain must make a
large number of improbable jumps through regions of low probability. To
properly weigh these modes, a Markov chain must alternate between them
frequently.  Two commonly used techniques to achieve this are \emph{parallel
tempering} (PT) and \emph{differential evolution}.

\subsection{Parallel Tempering}

\emph{Tempering} introduces a ``temperature''~$T$ into the likelihood function,
resulting in a modified posterior
\begin{equation}
  \label{eqn:PTposterior}
  p_T(\boldsymbol{\theta|d}) \propto
  \pi(\boldsymbol{\theta})\mathcal{L}(\boldsymbol{\theta})^{\frac{1}{T}} \;.
\end{equation}
Increasing temperatures above $T=1$ reduces the contrast of the likelihood
surface, shortening and broadening the peaks in the distribution and making
them easier to sample.  PT originates from Monte Carlo simulations in
condensed-matter physics, starting from Replica-Exchange Monte
Carlo~\cite{swendsen86} and then generalized to the full exchange of
``configurations''~\cite{parallelTempering}, cf.\ Ref.~\cite{earl05} for a
review.  It exploits the ``flattening'' of the distributions with increasing
temperature to construct an ensemble of tempered chains with temperatures
spanning $T=1$ to some maximum temperature $T_\text{max}$.  Chains at higher
temperatures are more likely to accept jumps to lower posterior values and
hence more likely to explore parameter space and move between isolated modes.
Regions of higher posterior value found by the high-temperature chains are then
passed down through the temperature ensemble via swaps between chains at
adjacent temperatures.  Such swaps are proposed periodically and accepted at a
rate $r_s = \text{min}(1, \omega_{ij})$, where
\begin{equation}
  \label{eqn:PTacceptance}
  \omega_{ij} =
  \left(\frac{\mathcal{L}(\boldsymbol{\theta_j})}
    {\mathcal{L}(\boldsymbol{\theta_i})}\right)^{\frac{1}{T_i}-\frac{1}{T_j}}
  \;,
\end{equation}
with $T_i < T_j$.  This technique greatly increases the probability of the
$T=1$ chain sampling between modes, but does so by creating many additional
chains whose samples are ultimately discarded, since they are not drawn from
the target posterior.  In our calculations, the temperatures~$T_i$ are
distributed logarithmically.  Every 100 iterations, swaps are proposed
sequentially between adjacent chains starting from the highest-temperature
pair.  All runs using the standard PT approach are done using 8 chains,
consistent with the analyses conducted during the last LVC science
run~\citep{S6PE}.

\subsection{Differential Evolution}

\emph{Differential evolution} attempts to solve the multimodal sampling
problem by leveraging information gained previously in the
run~\citep{differentialEvo}.  It does so by drawing two previous samples
$\boldsymbol{\theta_1}$ and $\boldsymbol{\theta_2}$ from the chain, and
proposing a new sample $\boldsymbol{\theta}'$ according to
\begin{equation}
  \label{eqn:DE}
  \boldsymbol{\theta}' = \boldsymbol{\theta} +
  \gamma(\boldsymbol{\theta_2}-\boldsymbol{\theta_1}) \;,
\end{equation}
where $\gamma$ is a free coefficient. 50\% of the time we use this as a
mode-hopping proposal, with $\gamma=1$.  In the case where
$\boldsymbol{\theta_1}$ and $\boldsymbol{\theta}$ are in the same mode, this
proposes a sample from the mode containing $\boldsymbol{\theta_2}$.  The other
50\% of the time we choose $\gamma$ uniformly between 0 and 1 to sample along
correlations.  This proves useful when linear correlations are encountered, but
performs poorly on non-linear correlations.

\subsection{Previous Implementation}
\label{subsec:previous}

The MCMC implementation used during the last LVC science run by
\emph{LALInference} employed a combination of PT and differential
evolution~\citep{PEmethods}.   Eight tempered chains were typically employed, with
computation time per chain ranging from several hours to 1--2 weeks, depending
on the waveform model used.  Although this approach proved effective at
sampling multimodal distributions, it required up to several thousand CPU hours
for a single run due to the number of samples collected at $T>1$ that did not
contribute to the estimation of the posterior.

\section{Parallel-Tempered Tuning}
\label{sec:newMCMC}

We proposed a pragmatic approach to address the high computation cost
associated with the conventional PT implementation.  PT is effective at
proposing jumps between isolated modes, but requires $n_\text{temps}-1$
additional likelihood evaluations for each sample in the $T=1$ chain, where
$n_\text{temps}$ is the number of parallel chains.  Differential evolution is a
computationally less expensive method to propose inter-modal jumps, but the
differential evolution buffer (i.e., sampling history) must first be filled
with samples across the posterior.  Even if the history of a chain represents a
perfect sampling of the posterior, there is only a probability $(n-1)/n^{2}$ of
drawing an inter-modal jump vector originating from the mode the chain is
currently in, for the case of a posterior with $n$ modes of equal weight.

\begin{figure}[ht!]
  \includegraphics[width=0.95\linewidth]{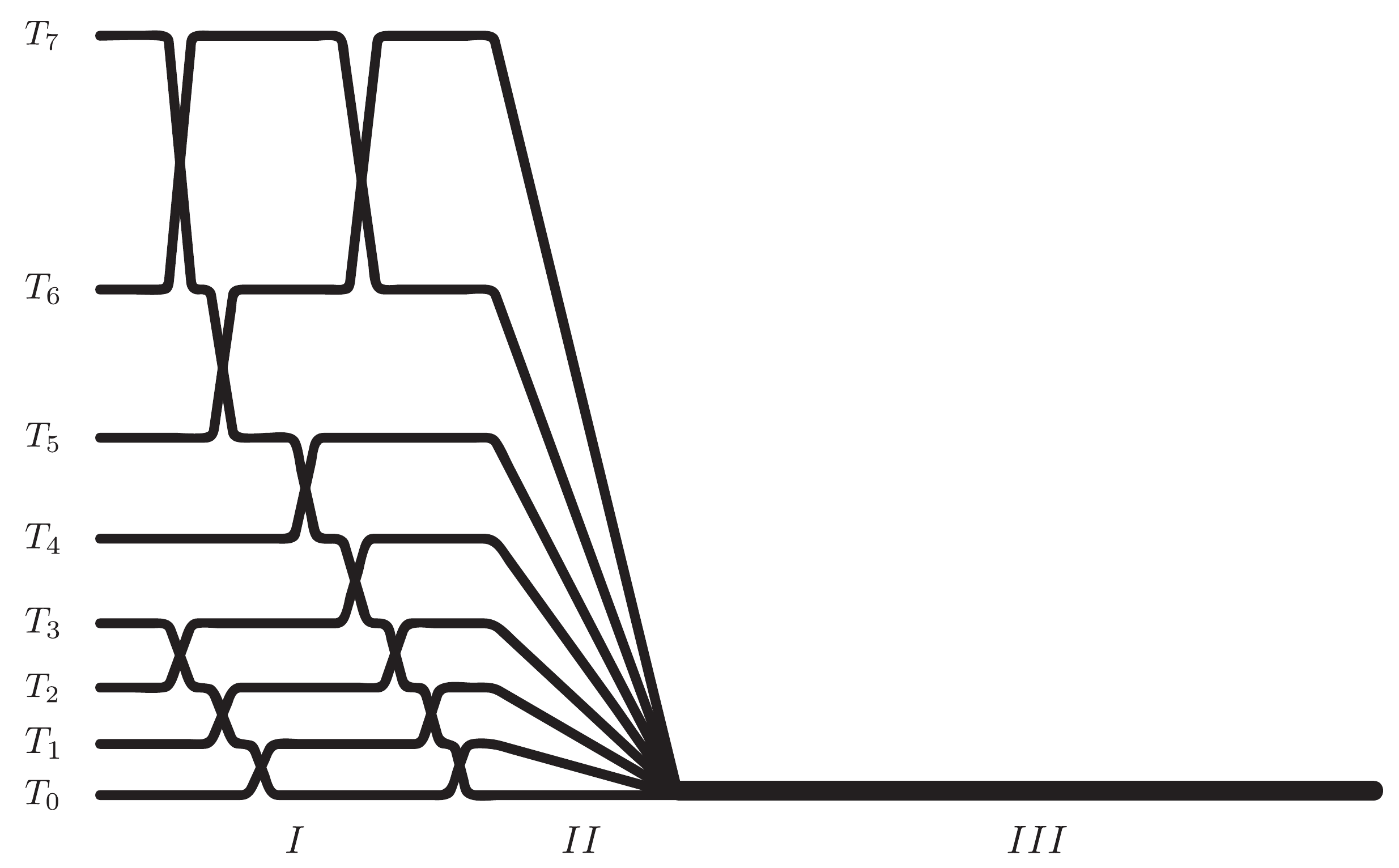}
  \caption{\label{fig:schematic} Schematic of the parallel-tempering tuned
      approach.  Each line represents a chain with a temperature increasing
      vertically.  Swaps in chain locations show when PT is in effect.  Phase I
      is the parallel-tempered burn-in which ends after $\mysim500$ effective
      samples, at which point the $T=1$ chain shares its differential evolution
      buffer with the other chains and the specialized proposal is tuned using
      its samples.  During phase II the $T>1$ chains are linearly annealed to
      $T=1$ over the course of $\mysim10$ ACTs.  Phase III produces all samples
      used to estimate the posterior, where chains sample independently using a
      jump proposal optimized to the target posterior.}
\end{figure}

To remedy this situation, we propose a new approach which uses parallel
tempering only during the burn-in phase.  The purpose of this short PT phase is
to allow the $T=1$ chain to collect samples from each of the isolated modes of
the posterior.  Once collected, these samples are used to produce a specialized
jump proposal that is tuned to the target posterior.  This new proposal
eliminates the need for PT chains, thus the $T>1$ chains can be
cooled to $T=1$, where they sample independently using the tuned proposal.
Figure~\ref{fig:schematic} shows a rough schematic of this PT-tuned approach.

\subsection{PT-tuned jump proposal}
\label{subsec:optics}

Central to this approach is a method of producing a proposal distribution from
the samples collected during the PT burn-in phase.  A kernel-density estimator
(KDE) is an obvious choice, as it produces a continuous distribution from a
sample set, and is trivial to draw samples from.  Given a set of samples
$\{\boldsymbol{x_1},\boldsymbol{x_2},...,\boldsymbol{x_n}\}$ drawn from the
target distribution $f$, its Gaussian KDE is given by
\begin{equation}
    \label{eqn:kde}
    \hat{f}_h(\boldsymbol{x})\propto\sum^n_{i=1}\text{exp}\left(\frac{-(\boldsymbol{x}-
            \boldsymbol{x_i})^T\cdot\boldsymbol{\Sigma^{-1}}\cdot(
            \boldsymbol{x}-\boldsymbol{x_i})}{2h^2}\right),
\end{equation}
where $\boldsymbol{\Sigma}$ is the covariance of the sample, and $h$ is the
bandwidth, which we have defined using Scott's Rule~\citep{scottsrule} to be
\begin{equation}
    \label{eqn:scotts}
    h = n^{-1/(d+4)},
\end{equation}
with $d$ the number of dimensions.  A sample is drawn from the estimated
distribution by first drawing a point $\boldsymbol{x_i}$ from the sample, then
drawing a point from a Gaussian centered on that point with covariance
$\boldsymbol{\Sigma}$.

However, KDEs tend to artificially broaden the modes of multimodal
distributions, which results in a poor estimate of the posterior and in low
proposal acceptance rates.  To avoid such ``over-smoothing'' we first cluster
the collected samples, identifying isolated areas of high posterior density and
effectively partitioning the parameter space into subspaces.  With each
partition containing a single mode of the posterior, an individual KDE can be
used to estimate the posterior in each partition with little over-smoothing.
These individual KDEs are weighted by the fraction of total samples contained
within the partition, then combined to produce a single estimate of the
posterior distribution across the full parameter space.

For the clustering step we have elected to partition the PT samples using
OPTICS (``Ordering Points To Identify the Clustering
Structure")~\citep{Ankerst:1999:OOP:304182.304187}, a density-based algorithm
designed to order a set of samples based on their density in parameter space.
From this ordering, a tree-based method~\citep{clusterTree} is used to extract
the clustering structure.  This approach does not require the number of
clusters to be known a priori, nor does it expect clusters to follow a
particular distribution.  The only input parameters required are the maximum
distance~$\varepsilon$ to check for nearest neighbors, and the minimum number
of points $N_\text{min}$ defining a cluster.  Once the clustering tree is
determined, each ``leaf'' is treated as a partition for which a KDE is
calculated.

\begin{figure}[ht!]
  \subfigure[]{
    \includegraphics[width=0.95\linewidth]{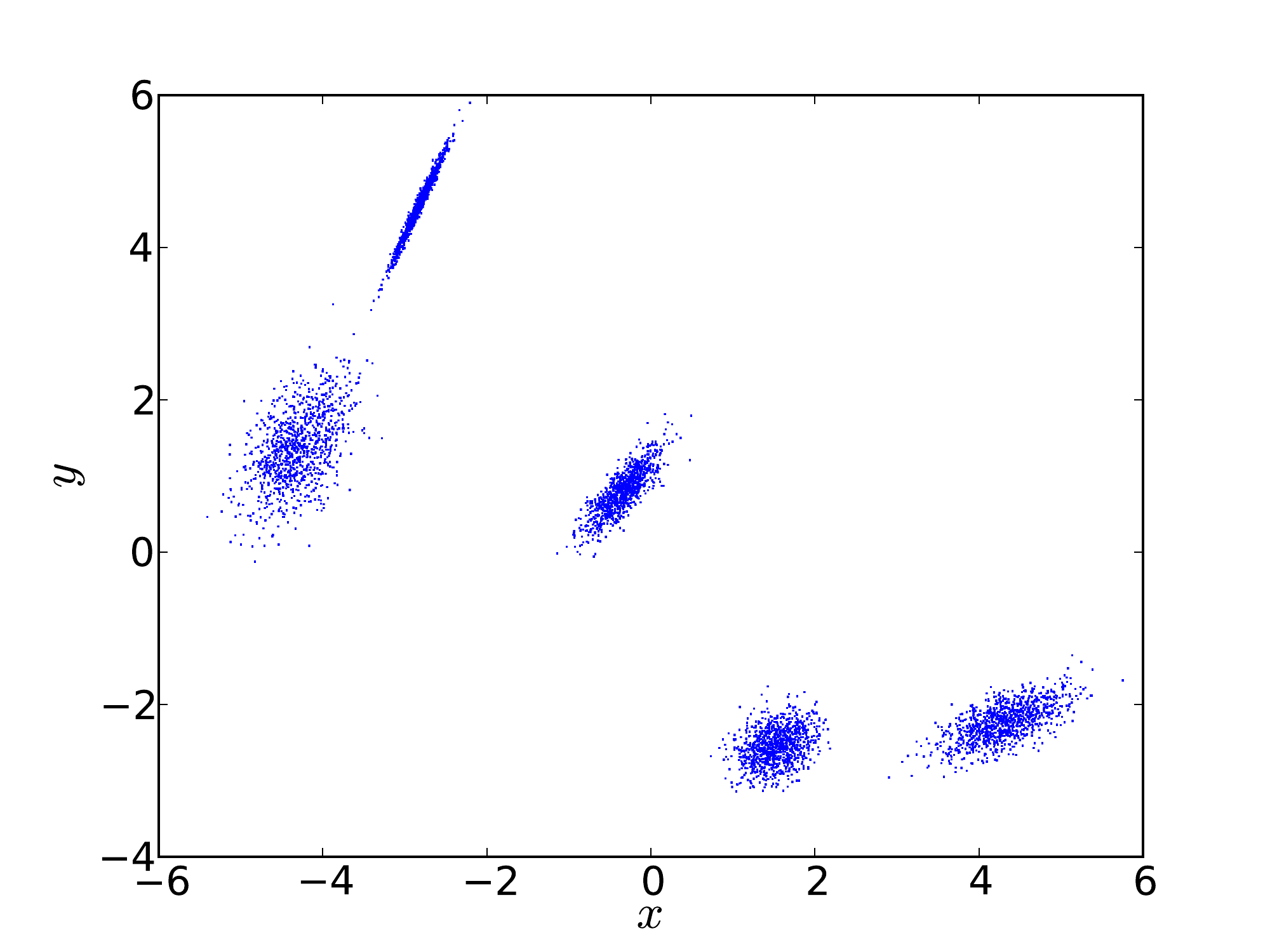}
    \label{subfig:samps}
  }

  \subfigure[]{
    \includegraphics[width=0.95\linewidth]{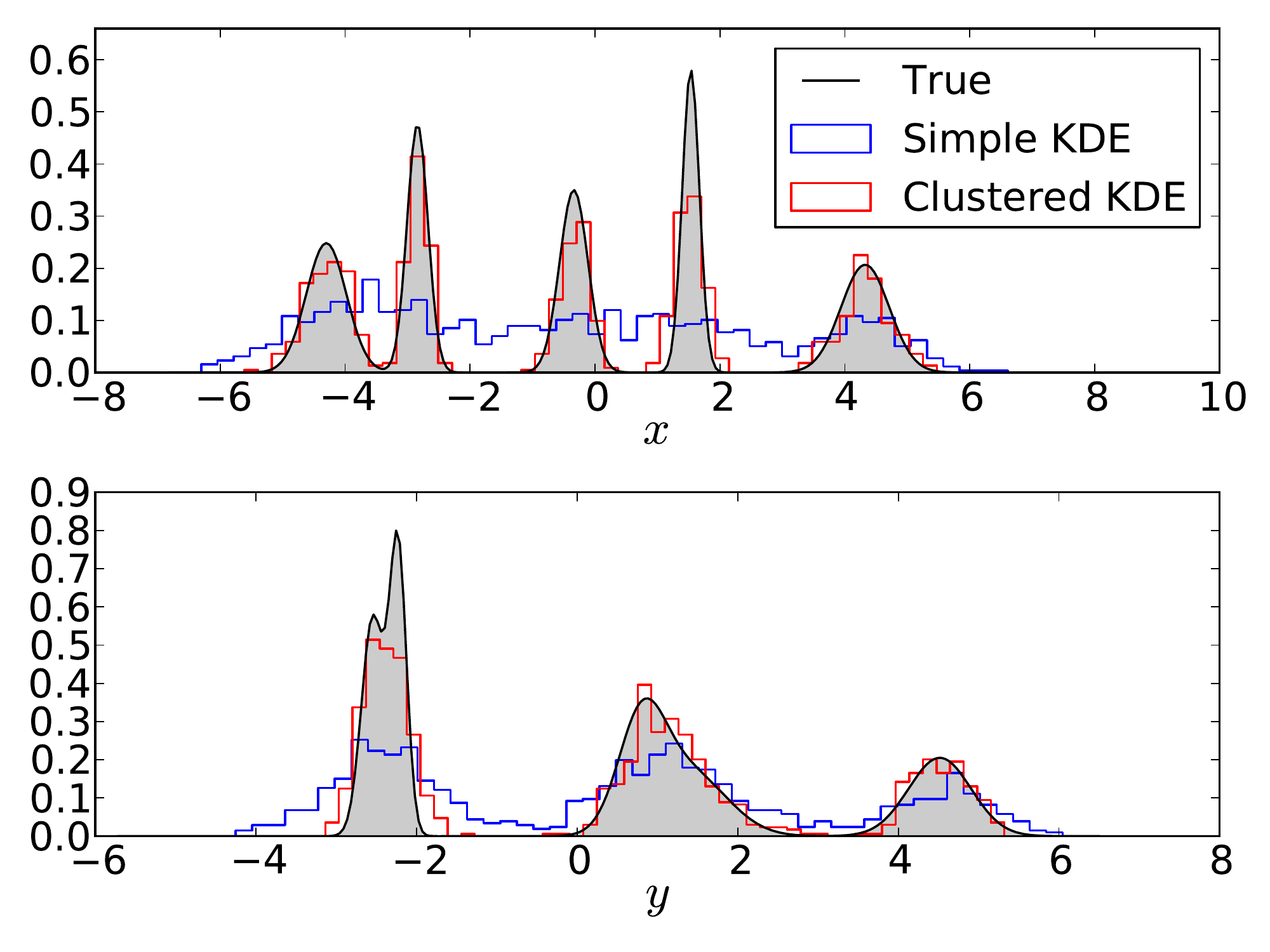}
    \label{subfig:proposals}
  }
  \caption{\label{fig:proposal_test} 
      An illustration of the clustered-KDE approach to estimating a
      distribution from a set of samples. \subref{subfig:samps} Samples
      drawn from a distribution composed of several 2D Gaussians with random
      sizes and orientations.  The distribution was then estimated from these
      samples using both the simple KDE and the clustered-KDE methods.
      \subref{subfig:proposals} Comparison of samples drawn from these
      estimates and the original set drawn from the true distribution.}
\end{figure}

A sample is generated from this proposal density by first drawing a leaf from
the tree, where leaf~$c$ is drawn with a probability
\begin{equation}
  \label{eqn:leafWeight}
  \gamma_c = \frac{N_c}{\sum_{\ell\in \mathcal{C}} N_\ell} \;,
\end{equation}
where $N_\ell$ is the number of samples in leaf~$\ell$ and $\mathcal{C}$ the
set of all leaves in the tree.  A sample is then drawn from the estimate of the
posterior in the leaf's subspace $p_c(\boldsymbol{\theta})$ as estimated by the
kernel-density estimator.

To illustrate the faithfulness of the clustered-KDE of a distribution over the
simple KDE, we combined several two-dimensional (2D) Gaussians with random
sizes and orientations.  Samples were drawn from this ``true'' distribution in
Fig.~\ref{subfig:samps}, which were then used to estimate the underlying
distribution using both methods.  Samples were then drawn from these estimates
and compared to the true distribution. Even in this simple 2D case the
clustered-KDE can be seen to be remarkably more faithful to the target
distribution (Fig.~\ref{subfig:proposals}), resulting in much higher acceptance
rates for the proposal.

To ensure that detailed balance is maintained, the forward and backward jump
probabilities must be computed to determine the acceptance
probability~\eqref{eqn:acceptance}.  In this case
$Q(\boldsymbol{\theta}'|\boldsymbol{\theta})$, the probability of proposing a
jump to $\boldsymbol{\theta}'$ from $\boldsymbol{\theta}$, is given by
\begin{equation}
  \label{eqn:jumpProb}
  Q(\boldsymbol{\theta}'|\boldsymbol{\theta}) = Q(\boldsymbol{\theta}')
  = \sum_{\ell\in \mathcal{C}} \gamma_\ell p_\ell(\boldsymbol{\theta}') \;.
\end{equation}
Since the jump proposal is independent of the chain's current location,
proposals are never correlated.  This reduces ACTs and
thereby increases the effective sample size for a chain of a given length.

Lastly, the KDE is able to accurately estimate distributions with modes of
arbitrary shape.  This makes the proposal efficient for proposing jumps along
non-linear correlations as well, addressing a shortcoming of differential
evolution.

\subsection{Annealing for Efficient Use of Chains}

During the parallel tempering phase, typically several hundred effective
samples are collected by the $T=1$ chain.  Once all modes in the posterior have
been sampled to some extent, the PT-tuned jump proposal and differential
evolution buffer will propose frequent inter-modal jumps.  This eliminates the
need for PT, allowing us to anneal all chains to $T=1$ where they independently
sample the target posterior distribution.  Here we have chosen the annealing
function of chain~$i$ to evolve linearly with iteration number from its
original temperature~$T_i$ to~$T=1$ over the course of $100~\ell_\text{PT}$
iterations, where $\ell_\text{PT}$ is the ACT of the $T=1$ chain during the
parallel tempering phase.  The cooling rate and precise mathematical form of
the cooling function were found to have no strong effect on the sampling
efficiency of this approach.

Once all chains have reached $T=1$, they no longer communicate.  From this
point onwards all chains are drawing samples from the target distribution, and
with the PT-tuned jump proposal are able to do so with shorter ACTs.  The jump
proposal set used for the final phase consists of 20\% PT-tuned proposals, 50\%
differential-evolution draws, 25\% Gaussian proposals, and 5\% proposals that
account for an exact degeneracy between $\phi_c$ and $\psi$.  Testing showed
this set of proposals to be effective for simulated GW data sets, however
extensive testing to find the optimal proposal set was outside the scope of
this work.

\section{Efficiency Tests}

To achieve the most efficient analysis we must minimize the number of
likelihood computations that are ultimately discarded.  This means minimizing
the length of chains with $T>1$, and minimizing the ACTs of chains sampling the
target posterior.

To compare the efficiency of the algorithms we will define an
\emph{effective sampling rate}, $r_\text{eff}$, given by
\begin{equation}
  \label{eqn:effSampRate}
  r_\text{eff} =
  \frac{\sum_{i=1}^{n_\text{chains}}N_{\text{eff},i}}
  {\sum_{i=1}^{n_\text{chains}}N_{\text{iter},i}} \;,
\end{equation}
where $N_{\text{eff},i}$ is the number of effective samples collected by chain
$i$, and $N_{\text{iter},i}$ the total number of likelihood calculations
performed for chain~$i$.  If we consider the entire ensemble of chains in a
run, the effective sampling rate is the number of uncorrelated samples divided
by the total number of likelihood computations.  For a run using only parallel
tempering, $N_{\text{eff},i}=0$ for $i>1$, since only the $T=1$ chain samples
from the target distribution.  For all analyses to follow, 12 chains were run
in parallel.  For the GW analyses, a maximum temperature $T_\text{max}$ was
chosen for each injection such that the maximum likelihood value was of order
10,
$\mathcal{L}_\text{max}$ followed
\begin{equation}
    \label{eqn:maxTemp}
    \frac{1}{T_\text{max}}\text{log}(\mathcal{L}_\text{max}) \mysim 10,
\end{equation}
where 10 was chosen to ensure the chain with the highest temperature would be
effectively sampling the prior distribution.

\subsection{Analytical Likelihoods}
\label{subsec:analytic}

To test the ability of the our approach to sample posteriors with correlated
parameters and multiple modes we tested it on three test distributions.  We
chose these distributions in a 15-dimensional space to emulate the
dimensionality of the parameter space that this proposal will ultimately need
to handle (i.e., spinning compact binaries).  The distributions used for testing
consisted of (i)~a multivariate Gaussian (200:1 ratio between largest and
smallest widths); (ii)~a bimodal distribution with two isolated multivariate
Gaussians of the same shape and orientation, separated by 8 standard
deviations; (iii)~a Rosenbrock
function~\citep{rosenbrock,generalizedRosenbrock}
\begin{equation}
  \label{eqn:rosenbrock}
  f(x_1,x_2,\ldots,x_d) 
  = \sum_{i=1}^{d-1}\left[(1-x_i)^2 + 100(x_{i+1} - x_i^2)^2\right].
\end{equation}
The performance of the proposed method is compared to that of the previous
implementation (Sec.~\ref{subsec:previous}), which only used parallel tempering
and differential evolution.  In all cases the chains ran until $\mysim10^3$
effective samples were collected, not including the burn-in.  With standard PT
this amounts to running all chains for $\mysim 1000$ effective samples
after burn-in, whereas for the new method each chain is run for about
$1000/n_\text{chains}$ effective samples after annealing.

The 1D marginalized posteriors recovered by both methods pass
one-sample Kolmogorov--Smirnov (K--S) tests against the analytical 1D functions
for the unimodal and bimodal multivariate Gaussian likelihoods, and two-sample
K--S tests against each other for the Rosenbrock likelihood.  Acceptance rates
for the PT-tuned proposal lied in the range $6$--$20\%$. This
demonstrates that the new proposals are able to generate successful jumps
around non-trivial likelihood surfaces without the use of parallel tempering.

\begin{table}[t]
  \centering
  \begin{tabular}{|l||r|c|r|c|r|}
    \hline
    & \multicolumn{2}{|c|}{Standard PT} & \multicolumn{2}{|c|}{PT-Tuned} & \\ \cline{2-5}
    Distribution & ACT  & $r_\text{eff}$     & ACT &
    $r_\text{eff}$     & $\frac{r_\text{eff,new}}{r_\text{eff,old}}$
    \\ \hline \hline
    Unimodal            & 280  & $3.3\times10^{-4}$ & 200 & $4.2\times10^{-4}$ & 1.26 \\ \hline
    Bimodal             & 850  & $1.3\times10^{-4}$ & 120 & $1.2\times10^{-3}$ & 9.02 \\ \hline
    Rosenbrock          & 3280 & $3.5\times10^{-5}$ & 470 & $3.4\times10^{-4}$ & 9.71 \\ \hline
  \end{tabular}
  \caption{Efficiency comparison between standard parallel tempering and the
      PT-tuned proposal for various test functions.  The autocorrelation times
      (ACTs) and effective sampling rates~$r_\text{eff}$ reported for standard PT
      are the median values from 10 runs with different random seeds.
      The PT-tuned ACTs are the median values from the 12 chains after
      burn-in.}
  \label{tab:analyticLikelihoods}
\end{table}

Table~\ref{tab:analyticLikelihoods} compares the ACTs and effective sampling
rates of the two approaches for the tested analytical likelihoods.  In all
cases the PT-tuned proposal produces chains with shorter ACTs, although the
improvement is minimal for the unimodal likelihood.  The simple structure of a
single Gaussian makes the $\mysim 500$ effective sample PT burn-in criterion
unnecessarily long, since no alternative modes need to be found or weighed.
This minimizes the gain in efficiency possible, since the majority of the
chain's total length (85\% in this case) is still done in the PT phase.  In
practice even standard PT is not needed to sample such a simple distribution,
however we include it here as a useful benchmark.  These comparisons are
sensitive to both the length of the burn-in and the total number of effective
samples being collected.  Since the burn-in procedure of the new method
contains both a parallel-tempering phase of hundreds of effective samples and
the annealing phase, it is typically much more expensive.  Thus, the reduction
in ACT from the PT-tuned proposal must be substantial enough to warrant the
burn-in.  Thus situations in which the posterior is highly structured are the
best candidates for improvement.  In addition, for long runs (i.e., large
numbers of effective samples), the cost of the burn-in period becomes less
important and even small improvements in ACTs can result in far fewer
likelihood computations over the course of the run.  We note that the new
approach provides more efficient sampling both due to a decrease in ACT and due
to the fact that all chains contribute to the posterior.

\subsection{Simulated GW Data}

To ensure that our findings are relevant to the analysis of GW signals, we
performed additional comparisons using simulated Gaussian detector noise
containing simulated gravitational wave signals.  GW signals were generated
using the TaylorF2 template family~\citep{taylorf2}.  For this work we have
only included non-spinning compact binary mergers, restricting the parameter
space to nine dimensions. Ten different signals were tested with total masses
ranging from 1.4 to 12.8 $\text{M}_\odot$, and network signal-to-noise ratios
between 12.5 and 63.4.  Again, we found that comparisons of the estimated
posteriors pass the two-sided K--S test and are consistent with the injected
signal.  Acceptance rates for the PT-tuned proposal fell in the range of
$\mysim0.2$--$18\%$.  The $0.2\%$ acceptance rate was an outlier, with typical
acceptance rates on the upper end of this range.  However, even the low
acceptance rate did not hinder the run enough for it to be outperformed by the
standard PT approach.

\begin{table}[t]
  \centering
  \begin{tabular}{|c||r|c|r|c|r|}
    \hline
     & \multicolumn{2}{|c|}{Standard PT} & \multicolumn{2}{|c|}{PT-Tuned} & \\ \cline{2-5}
    Event & ACT & $r_\text{eff}$ & ACT & $r_\text{eff}$ & $\frac{r_\text{eff,new}}{r_\text{eff,old}}$ \\ \hline \hline
    1 & 1300 & $8.5\times10^{-5}$ & 1040 & $1.2\times10^{-4}$ & 1.4 \\ \hline
    2 & 2700 & $4.6\times10^{-5}$ & 190  & $5.8\times10^{-4}$ & 13 \\ \hline
    3 & 2160 & $5.2\times10^{-5}$ & 340  & $3.1\times10^{-4}$ & 5.9 \\ \hline
    4 & 1440 & $7.6\times10^{-5}$ & 430  & $3.2\times10^{-4}$ & 4.2 \\ \hline
    5 & 4220 & $2.8\times10^{-5}$ & 5500 & $9.4\times10^{-5}$ & 3.3 \\ \hline
    6 & 840  & $1.3\times10^{-4}$ & 270  & $2.4\times10^{-4}$ & 1.9 \\ \hline
    7 & 1540 & $7.8\times10^{-5}$ & 2030 & $9.8\times10^{-5}$ & 1.3 \\ \hline
    8 & 560  & $1.8\times10^{-4}$ & 200  & $2.7\times10^{-4}$ & 1.5 \\ \hline
    9 & 1460 & $7.9\times10^{-5}$ & 300  & $3.6\times10^{-4}$ & 4.6 \\ \hline
    10 & 960  & $7.9\times10^{-5}$ & 310  & $3.0\times10^{-4}$ & 3.9 \\ \hline
  \end{tabular}
  \caption{Efficiency comparison between standard parallel tempering and the
      PT-tuned approach for 10 randomly selected simulated non-spinning
      gravitational-wave signals.  The values reported for standard parallel
      tempering are collected from one run of 8 parallel chains.  The
      autocorrelation times for the new method are the median values from the
      12 chains after burn-in.}
  \label{tab:skyLocTests}
\end{table}

Table~\ref{tab:skyLocTests} compares the ACTs and effective sampling rates of
the two methods for 10 randomly selected non-spinning simulated gravitational
wave signals.  The PT-tuned approach shows improved efficiencies over standard
PT for all ``events,'' but unlike our findings for the analytical likelihoods,
here the ACTs of the chains are not always smaller with the PT-tuned proposal.
Nevertheless, in these cases the cost due to longer ACTs is still compensated
for by having all chains contribute to the effective sampling.  The large range
in efficiency improvement, from factors of 1.3 to 13 in these tests, can be
attributed to several factors.  The complexity of the posterior depends
strongly on the injection parameters, ranging from unimodal with little
correlation (e.g., event~1) to multimodal and highly correlated (e.g.,
event~2).  This strongly affects the efficiency gains that can be obtained with
the new approach, as was also shown by the analytical tests in
Sec.~\ref{subsec:analytic}.  Moreover, for some runs the posterior was poorly
estimated when constructing the jump proposal, a natural consequence of
estimating the posterior before it has been exhaustively sampled. This is the
cause for the small efficiency improvement seen for event~7, which had proposal
acceptance rates of $\mysim0.2$--$0.3\%$.  Despite these low acceptance rates,
the chains were still able to sample the whole posterior, and do so more
efficiently than standard PT.

\section{Summary}

We have presented an alternative implementation to standard parallel tempering,
designed to minimize the time spent parallel tempering to avoid likelihood
computations not contributing to the estimation of the posterior.  By parallel
tempering only long enough for the $T=1$ chain to identify the modes of the
posterior, a jump proposal can be tuned to the target posterior.  These tuned
jump proposals allow sampling of multiple modes and/or along non-linear
correlations without the continued use of parallel tempering while also
reducing ACTs.  These benefits come with the trade-off of a more expensive
burn-in process than traditional parallel tempering. However we find that for,
highly structured (i.e., non-Gaussian) likelihood surfaces the proposed
approach proves to be worth this cost. The gains in efficiency will increase
with the complexity of the posterior, making us optimistic that the more
complex posteriors encountered in the spinning parameter space of compact
binary mergers will see even larger increases in sampling efficiency.  Such a
study will be the subject of future work.

\begin{acknowledgments}
  This work is supported by the National Science Foundation under Grants
  PHY-0969820 for VK and EL, and DGE-0824162 for BF. VK also thanks the Aspen
  Center for Physics for its hospitality while she worked on this project.
\end{acknowledgments}

\bibliography{paper}

\end{document}